# Cold Pools Reduce the Impacts of Deforestation on Convection Initiation

Nicholas M. Falk[1], Gabrielle R. Leung[1], Leah D. Grant[1], and Susan C. van den Heever[1]

[1] Department of Atmospheric Science, Colorado State University, Fort Collins, CO, USA

Corresponding author: Nicholas M. Falk (nick.falk@colostate.edu)

Co-Author emails:

- Gabrielle.Leung@colostate.edu
- Leah.Grant@colostate.edu
- Sue.vandenHeever@colostate.edu

## Acknowledgments

Funding was provided by NSF Grant AGS-2029611. We would like to acknowledge high-performance computing support from the Derecho system (doi:10.5065/qx9a-pg09) provided by the NSF National Center for Atmospheric Research (NCAR), sponsored by the National Science Foundation. We thank Dr. Aryeh Drager for helpful discussions around this work.

## Conflict of Interest Statement

The authors declare no conflicts of interest.

## Data Availability Statement

Model source code, name lists, and analysis scripts are available at: https://doi.org/10.5281/zenodo.17654571

**Abstract**

The individual and synergistic impacts of cold pools and land surface heterogeneity on convection initiation are investigated. Idealized large eddy simulations of deep convection over the Amazon rainforest are conducted. Simulations test realistic and homogenized vegetation, along with realistic and suppressed low-level evaporation which eliminates cold pools. Updrafts are tracked to determine convection initiation locations. The aggregation of initiation locations is quantified with an organization index. Initiation locations are randomly distributed over homogeneous vegetation, with or without cold pools, demonstrating that cold pools have minimal impacts on initiation locations over homogeneous vegetation. With realistic vegetation, convection initiation is more frequent over forested than deforested areas due to favorable thermodynamics. Heterogeneous vegetation effectively aggregates initiation locations and precipitation compared to homogeneous vegetation, whereas cold pools disaggregate initiation locations and precipitation by propagating into deforested regions and initiating convection. Thus, cold pools partially counteract the effects of anthropogenically driven land surface heterogeneity.

**1 Introduction**

Anthropogenic deforestation in rainforests such as the Amazon creates a heterogeneous land surface. Such heterogeneities can drive mesoscale circulations thereby impacting the distribution of convective clouds and rainfall (Chagnon et al., 2004; Durieux et al., 2003; Paccini & Schiro, 2025; Souza et al., 2000; Wang et al., 2009). When land surfaces with different properties are adjacent on the mesoscale, studies have shown that moist convection is usually favored over warmer, drier surfaces compared to nearby cooler, moister surfaces (Chagnon et al., 2004; Mascart et al., 1991; Rabin et al., 1990; Taylor et al., 2012). Enhanced convective activity over warm, dry surfaces has been attributed to mesoscale solenoidal circulations driven by surface heterogeneity (hereafter simply solenoidal circulations). These circulations flow at low levels from cool, moist regions into warm, dry regions, ascend in warm, dry regions, and then return towards cool, moist regions at higher levels. Ascent from solenoidal circulations can initiate convection in warm, dry regions (Branch & Wulfmeyer, 2019; F. Chen & Avissar, 1994; W. Y. Y. Cheng & Cotton, 2004; Cioni & Hohenegger, 2018; Hong et al., 1995; Mascart et al., 1991). Vegetation-induced solenoidal circulations in tropical continental environments can initiate convection at their edges both by lifting air and by creating a favorable thermodynamic environment (Garcia-Carreras et al., 2011; Taylor et al., 2011). Collisions between solenoidal circulations can also initiate convection (Lynn et al., 1998; Rieck et al., 2014).

Similar to heterogeneous land surfaces, convective cold pools create regions that are favorable for new storm initiation and regions where convection initiation is suppressed. Moist convection tends to be inhibited in the interiors of cold pools (Byers & Braham, 1949; Drager & van den Heever, 2017). On the other hand, cold pools initiate new storms at their edges through the same mechanisms as solenoidal circulations: mechanical lifting, creating favorable thermodynamics, and collisions (Purdom, 1976; Tompkins, 2001; Torri et al., 2015; Weaver &

Nelson, 1982). These similarities suggest that convection initiation may be affected by the interaction between cold pools and solenoidal circulations.

Cold pools and solenoidal circulations can interact in important ways. Cold pools can accelerate or decelerate solenoidal circulations by altering the temperature gradient across the circulation. (Ascher et al., 2024; Harvey et al., 2022; Rieck et al., 2015). Chen et al. (2020) and Ascher et al. (2024) both found that convection which initially forms due to land surface heterogeneities can produce cold pools that subsequently initiate new convection in areas where convection was previously absent. Despite this previous research, the impacts of surface heterogeneity and cold pools on convection initiation have not been systematically disentangled, and the misrepresentation of cold pools and solenoidal circulations is a known source of error in forecast and climate models (Birch et al., 2015; Garcia-Carreras et al., 2013; Rooney et al., 2022; Yang et al., 2021). Studies which aim to understand the impacts of cold pools often conduct mechanism-denial experiments to eliminate cold pools through changes to evaporation (Crook & Moncrieff, 1988; Grant et al., 2018, 2020; Jeevanjee & Romps, 2013; Khairoutdinov & Randall, 2006), but such mechanism-denial experiments are usually conducted over homogeneous surfaces. Recently, Maybee et al. (2025) conducted a mechanism-denial experiment to eliminate cold pools in Sahelian case-study simulations with heterogeneous land surfaces, finding that cold pools were important for the structure of Sahelian MCSs but not for their initiation or maintenance.

The objective of this study is to address the following science question: What are the individual and synergistic impacts of surface heterogeneity/solenoidal circulations and cold pools on convective initiation? We run a set of large eddy simulations (LES) designed to directly address this question. Realistic and weakened cold pools are tested along with realistic and homogenized vegetation. Convective updrafts are tracked to identify where they initiate. Factor separation (Stein & Alpert, 1993) is used to disentangle the individual and synergistic impacts of both cold pools and surface heterogeneity on convection initiation.

## 2 Methods

2.1 Large Eddy Simulations

Large eddy simulations are conducted using the Regional Atmospheric Modeling System (RAMS) version 6.3.04 (Cotton et al., 2003; van den Heever et al., 2023; Pielke et al., 1992; Saleeby & van den Heever, 2013). Simulations are idealized to isolate processes of interest, except that realistic vegetation data are used to accurately capture the heterogeneity patterns created by deforestation. A horizontal grid spacing of 150 m is used with a timestep of 0.75 s, while the vertical grid spacing starts at 50 m near the surface and is stretched to 300 m. Such grid spacings are necessary to accurately represent cold pools and their interactions with the land surface (Fiévet et al., 2023; Grant & van den Heever, 2016; Hirt et al., 2020; Straka et al., 1993). The model domain is 150 x 150 x 24.2 km, is centered at 10.70 S, 62.05 W, and contains the site where the initial sounding was launched from (Figure 1a, where "evergreen broadleaf tree" is rainforest). Periodic lateral boundaries are used. The Land-Ecosystem-Atmosphere-

Feedback version 3 (LEAF-3) (Walko et al., 2000) model simulates two-way land-atmosphere interactions. Radiative tendencies are updated every 5 min using Radiative Transfer for Energetics+Rapid Radiative Transfer Model for General circulation model applications - Parallel (RTE+RRTMGP) (Pincus et al., 2019). Cloud microphysics are represented using the RAMS double-moment microphysics scheme (Saleeby & van den Heever, 2013).

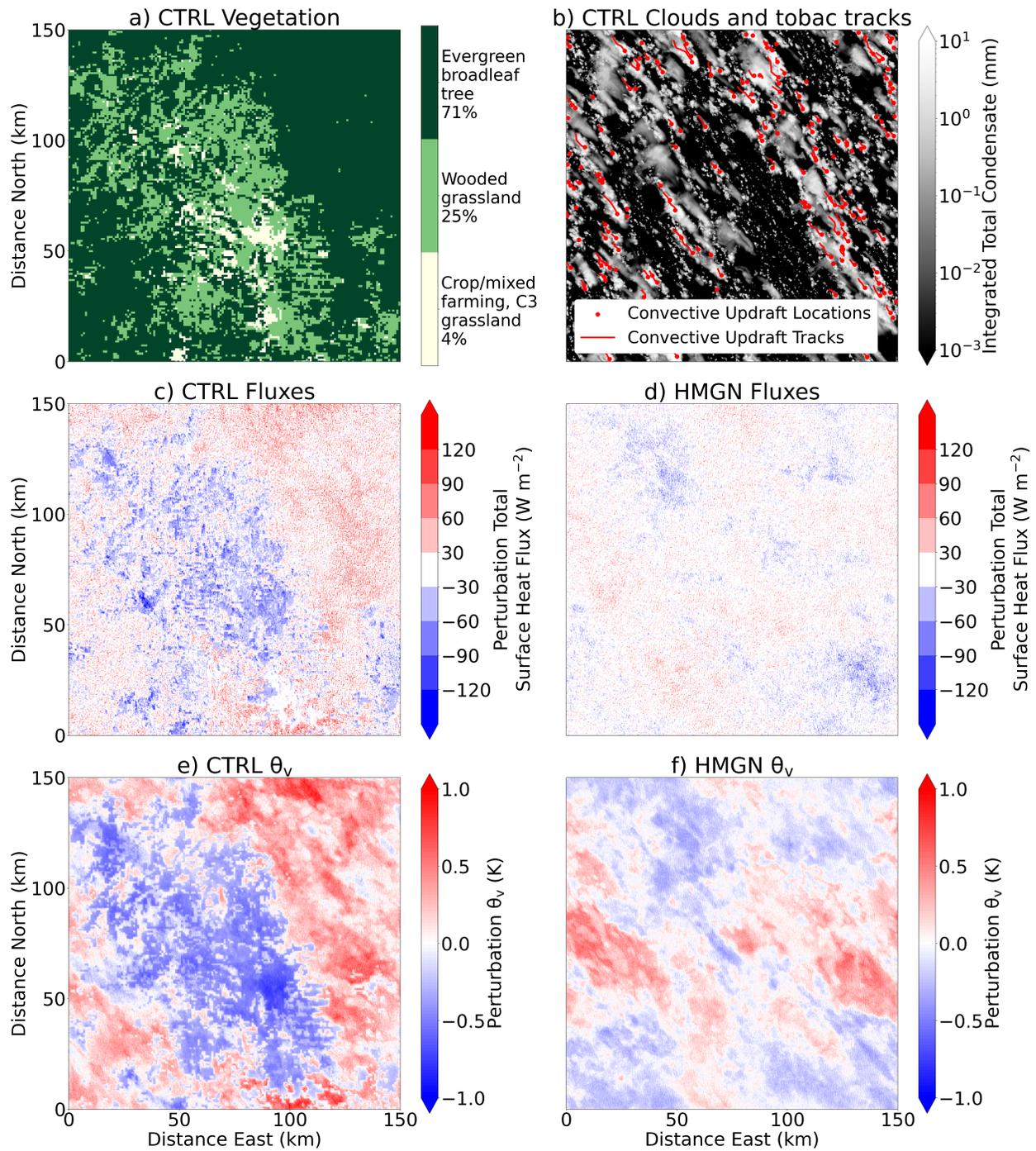

Figure 1: **Overview of the basic simulation characteristics**. (a) Vegetation class in CTRL (shading) with percent of the domain covered by each class indicated in the colorbar. (b)

Vertically integrated total condensate (shading), *tobac*-identified convective updraft locations (red points), and convective updraft tracks (red lines) in CTRL at 1200 LT on day one. (c) Perturbation total surface heat flux (shading) and (e) 25 m AGL perturbation virtual potential temperature (shading) in CTRL. Perturbations are calculated from the domain- and time-mean at 0930 LT across all 5 simulated days. (d,f) As in (c,e) but for HMGN.

The atmosphere is initialized using the 23 February 1999 TRMM-LBA sounding presented in Grabowski et al. (2006) (Figure S1). This sounding is chosen because it has been commonly used in studies of Amazon convection and in mechanism-denial studies of cold pools (Grabowski et al., 2006; Khairoutdinov & Randall, 2006; Kurowski et al., 2018). Simulations start on 23 February 1999 at 0730 LT (UTC = LT + 4). They are integrated for 5 days to capture multiple diurnal cycles and ensure that large numbers of convective updrafts are simulated, thus generating a statistically robust sample of convection initiation events. Sandy loam soil, an initial soil temperature of 24.8 °C, and an initial soil moisture 0.24 $m^3$ $m^{-3}$ are chosen based on contemporaneous observations shown in Alvalá et al. (2002). The soil extends 2 m below the surface and is broken into 26 levels. Note that all atmospheric and land surface fields are initially horizontally homogeneous in all simulations, other than the vegetation in certain simulations (as described below). Initial random potential temperature perturbations of amplitude 0.1 K in the lowest 500 m are used to facilitate boundary layer development.

Four simulations are conducted to assess the individual and synergistic impacts of land surface heterogeneity and cold pools on convection initiation. In the control simulation (CTRL), realistic vegetation and realistic evaporation are used, thereby representing realistic land surface and cold pool processes. Figure 1a shows the vegetation in CTRL. Note that the vegetation fraction does not vary within each of the three vegetation classes in CTRL. In the homogeneous simulation (HMGN), vegetation properties at each grid point are set to the vegetation-area-weighted mean values of properties in CTRL. Thus, CTRL and HMGN have the same domain-mean vegetation properties, but CTRL has heterogeneously distributed vegetation while HMGN does not. This homogenization only applies to properties which are fixed in time, so prognostic quantities (e.g. soil and canopy moisture/temperature, surface fluxes) are allowed to evolve freely in all simulations. Thus, surface heterogeneities in quantities like soil moisture can develop over time in HMGN. In the evaporation-off simulation (EOFF), evaporation of rain and drizzle are disabled in the lowest 2 km, following Grant et al. (2018). Only the evaporation of falling rain and drizzle are disabled, so other cloud microphysical processes, such as the evaporation of cloud water due to entrainment, remain unaltered. Analysis of density potential temperature at the lowest model level above ground shows that cold pools are almost entirely eliminated in EOFF (Figure S2). Finally, the homogeneous vegetation and evaporation off simulation (HMGN_EOFF) simultaneously applies the homogeneous vegetation of HMGN and the disabled evaporation of EOFF to assess synergistic interactions between these processes.

## 2.2 Tracking

We use a tracking algorithm to identify where convective updrafts initiate. By tracking convective updrafts in time, we can distinguish the locations where convective updrafts initiate from the locations where convective updrafts subsequently propagate. We focus our analysis only on the initiation locations of convective updrafts to prevent double-counting. The Tracking and Object-Based Analysis of Clouds (*tobac*) algorithm v1.5 (Heikenfeld et al., 2019; Sokolowsky et al., 2024) is used to track convective updrafts in four dimensions (zonal, meridional, vertical, and time). A vertical velocity threshold of 3 m s$^{-1}$ and a volume threshold of at least 64 contiguous grid points in any spatial dimension is used to identify convective updrafts. Updrafts tracked for less than 15 min are excluded from further analysis, as updrafts of such durations are likely shallow convection. Updrafts tracked for more than 15 min are referred to as convective updrafts (e.g. Figure 1b, red points and lines). The selected *tobac* thresholds were subjectively determined after trial-and-error, but results are not qualitatively sensitive to the thresholds chosen.

## 2.3 Quantifying Aggregation

We use the organization index ($I_{org}$) (Tompkins & Semie, 2017) to quantify the aggregation of convection initiation locations. For a given simulation, the nearest neighbor distances between all convection initiation locations are determined. $I_{org}$ is calculated by plotting the cumulative density function of these nearest neighbor distances (nearest neighbor cumulative density function, or NNCDF) against a NNCDF of random uniform points. The area under this curve is defined as $I_{org}$. An $I_{org}$ value of ~0.5 indicates the initiation locations are randomly distributed, while values above (below) 0.5 indicate the initiation locations are aggregated (regularly distributed). $I_{org}$ is a dimensionless quantity. All convection initiation locations over the five simulated days are used when calculating $I_{org}$ for a given simulation. We therefore quantify the degree to which the locations where convective updrafts initiate are aggregated. Our use of $I_{org}$ is distinct from past studies that have used $I_{org}$ to quantify the organization of convection at a given snapshot in time (W.-Y. Cheng et al., 2018; Pscheidt et al., 2019; Seifert & Heus, 2013). For each simulation, a distribution of $I_{org}$ values is calculated by generating 1000 random uniform NNCDFs and thus recalculating $I_{org}$ 1000 times. For a given simulation X, $I_{org,mean,X}$ represents the mean value of $I_{org,mean,X}$ across these 1000 recalculations.

## 2.4 Factor Separation

Factor separation (Stein & Alpert, 1993) is applied to understand the individual and synergistic impacts of heterogeneous vegetation and cold pools on patterns of convection initiation. The individual impact of heterogeneous vegetation ($F_{VEG}$) on $I_{org}$ is calculated as:

$$F_{VEG} = I_{org,mean,EOFF} - I_{org,mean,HMGN\_EOFF} \qquad (1),$$

the individual impact of cold pools ($F_{CP}$) is calculated as:

$$F_{CP} = I_{org,mean,HMGN} - I_{org,mean,HMGN\_EOFF} \qquad (2),$$

and the impact of synergistic interactions between heterogeneous vegetation and cold pools ($F_{VEG\_CP}$) is calculated as:

$$F_{VEG\_CP} = I_{org,mean,CTRL} - I_{org,mean,HMGN\_EOFF} - F_{VEG} - F_{CP} = I_{org,mean,CTRL} - (I_{org,mean,HMGN} + I_{org,mean,EOFF}) + I_{org,mean,HMGN\_EOFF} \quad (3).$$

See Appendix S1 for a detailed explanation of the factor separation method.

## 3 Results

Vegetation heterogeneity clearly impacts surface heat fluxes and low-level virtual potential temperature. The total surface heat flux (sum of sensible and latent heat fluxes) at 0930 LT, before deep convection onset, is greater over the forest than the deforested region in CTRL (Figure 1c). Surface heat fluxes are greater over the forest because the surface roughness length is greater and the albedo is smaller over the forested than the deforested region (Table S1). A reduced albedo implies the forest absorbs more energy from solar radiation, while greater roughness lengths mean the forest more efficiently transfers this energy to the atmosphere through turbulent processes. Accordingly, at 0930 LT, the 25 m virtual potential temperature in CTRL is greater over the forest than over the deforested region (Figure 1e). The homogeneous vegetation in HMGN causes perturbations in surface heat fluxes and low-level virtual potential temperature that are smaller in magnitude than in CTRL (compare Figure 1c-f). Perturbations of pre-convective surface heat fluxes and virtual potential temperature in EOFF (HMGN_EOFF) are qualitatively similar to those for CTRL (HMGN) (not shown).

Convective updrafts in CTRL and EOFF initiate less frequently over the deforested region than over the forest (Figure 2a,c). The lack of convective updrafts in the deforested region can be explained by the reduced virtual potential temperature, and hence the air being less buoyant, compared to the forest (Figure 1a,e). At low-levels, the forest is warmer but slightly drier than the deforested region (Figure S3), a response to deforestation which has been shown to occur in tropical regions with high soil moisture (van der Molen et al., 2006). The forest is drier than the deforested region at low-levels partially due to enhanced vertical mixing of moisture over the forest (Figure S3d). The locations of convection initiation before 1200 LT in CTRL and EOFF demonstrate the effects of vegetation-driven solenoidal circulations and cold pools on convective updraft initiation (Figure 2b,d). During this time period, convection initiates most frequently near the forested-deforested boundary. These patterns are consistent with solenoidal circulations beginning to propagate from the deforested region into the forest and initiating convection. As the day progresses, these solenoidal circulations propagate further into the forested region and continue to initiate convection (not shown). Therefore, while solenoidal circulations are evident in the simulations, their effects are muted in Figure 2a,c because the entire diurnal cycle is shown and the effects of the solenoidal circulations are spread across the forest.

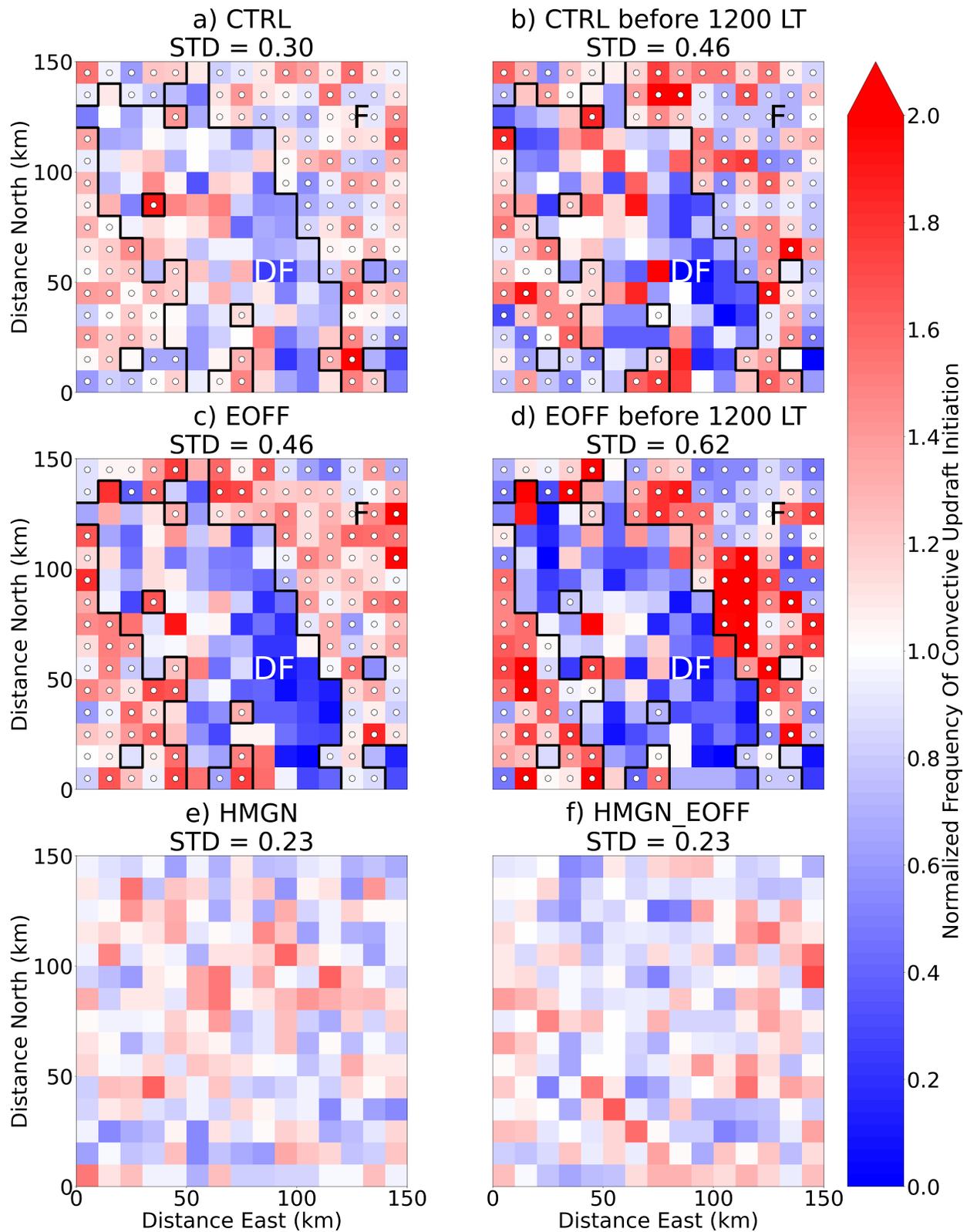

Figure 2: **Convective updrafts initiate more frequently over forested than deforested areas, especially when cold pools are not present.** (a,c,e,f) Heatmaps of convection initiation

locations for CTRL, EOFF, HMGN, and HMGN_EOFF, respectively; values are normalized such that 1.0 is the domain mean for each panel, and a value of 2.0 represents twice the number of initiating convective updrafts compared to 1.0. (b,d) as in (a,c) but only for convective updrafts which initiate between 0000 LT and 1155 LT. White dots indicate bins with more forest cover than the domain mean (71%). Black contours separate forested and deforested areas. Labels F (DF) also indicate forest (deforested region). The standard deviation of plotted values is included in the title of each panel.

Convective updrafts in HMGN and HMGN_EOFF initiate more uniformly across the domain than convective updrafts in CTRL and EOFF (Figure 2a,c,e,f). Virtual potential temperature is more uniform across the domain with homogeneous vegetation compared to heterogeneous vegetation (Figure 1e,f). Therefore, convection initiation is not favored nor disfavored in specific regions, nor can solenoidal circulations form in the homogeneous vegetation simulations.

Cold pools also influence where convection initiates. Recall the only difference between EOFF and CTRL is the suppressed evaporation of precipitation in the lowest 2 km, and therefore lack of cold pools, in EOFF. Convective updrafts in EOFF initiate less frequently over the deforested region than convective updrafts in CTRL (compare Figures 2a,c). This suggests that cold pools in CTRL propagate from the forest into the deforested region and initiate convective updrafts over the deforested region, despite the less favorable conditions for convection initiation compared to the forest. On the other hand, in EOFF, very few convective updrafts initiate in the deforested region due to the lack of cold pool-induced initiation. To further support this argument, we return to Figure 2b. The time period shown captures the first set of convective updrafts on each day in CTRL which initiate before strong cold pools form, thus partially removing the influence of cold pools. Early convection initiation frequency in CTRL is suppressed in the deforested region relative to later time periods (compare Figures 2a,b) and even approaches the magnitude of suppression in EOFF over the full diurnal cycle (compare Figures 2b,c), further demonstrating that cold pools are largely responsible for the initiation of convective updrafts in the deforested region of CTRL.

To investigate the mechanisms driving convective initiation, we create composites of low-level density potential temperature, water vapor mixing ratio, and vertical velocity around each convective updraft initiation. We further categorize each convective updraft initiation based on whether it occurs near no, one, or multiple density currents (cold pools and/or solenoidal circulations) (Figure 3). Composites are rotated so that density currents are on the left side of each panel in Figure 3, and flipped so that colliding density currents are on the bottom right of each panel in Figure 3c,f,i,l. Appendix S2 describes the procedure for categorization and compositing in detail. Cold pools play a major role in initiating convection, as density currents initiate ~58% of convective updrafts in HMGN (Figure 3e-f), where cold pools are present, but only ~9% of convective updrafts in HMGN_EOFF, where cold pools are absent (Figure 3k-l). Solenoidal circulations play a secondary role, as ~14% of convective updrafts are initiated by one or more density currents in EOFF (Figure 3h-i), compared to ~9% in HMGN_EOFF (Figure 3k-l). Regardless of the simulation, density currents are associated with enhanced water vapor mixing ratio and vertical velocity at their edges (Figure 3), which previous studies have shown can lead to the initiation of new convection (Garcia-Carreras et al., 2011; Purdom, 1976; Tompkins, 2001). Note that some density currents are still present in HMGN_EOFF (Figure 3j-l) as cold pools are weakened, but not completely eliminated, in this simulation and in EOFF.

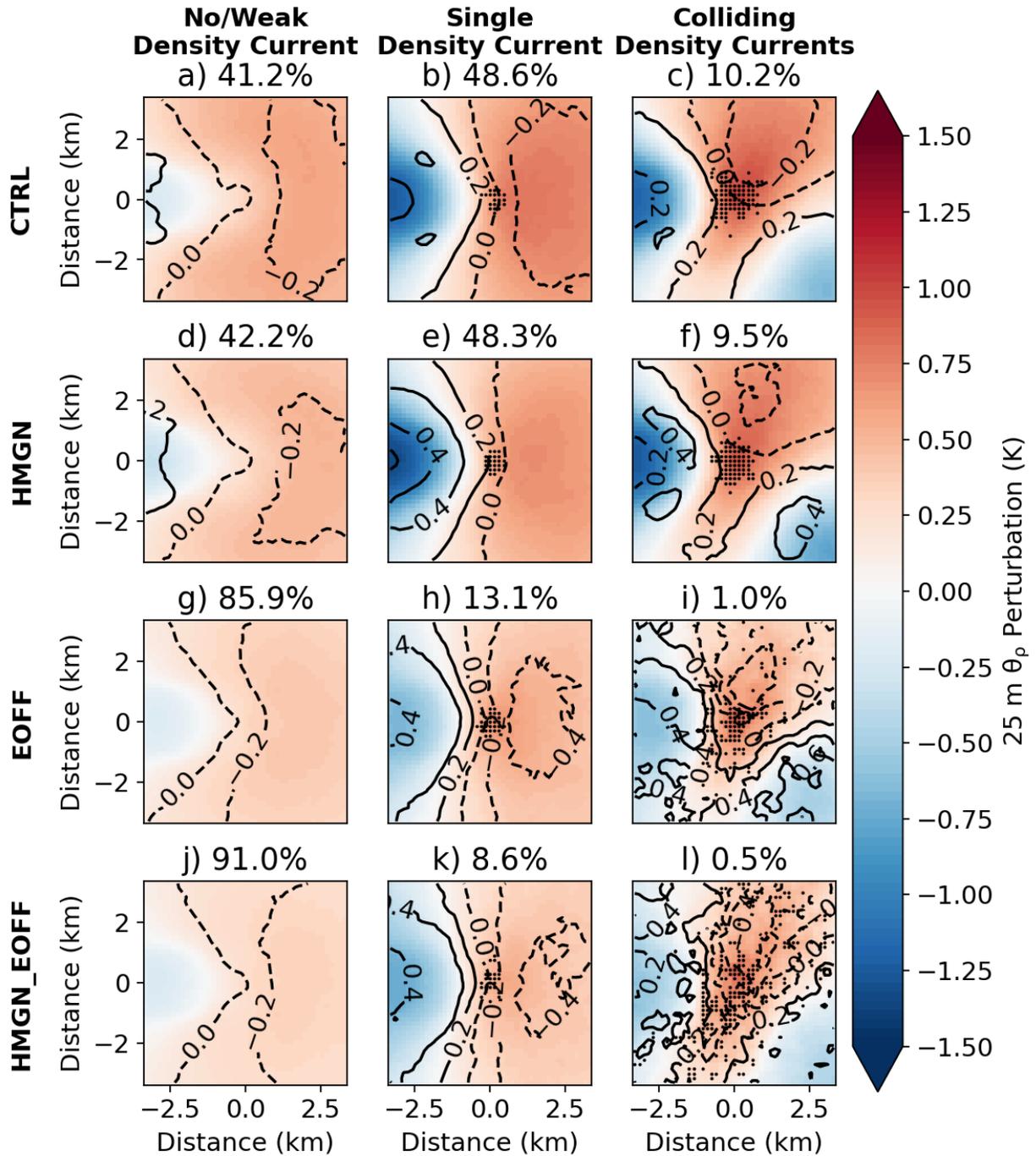

Figure 3: **Density currents initiate convection by enhancing water vapor mixing ratio and vertical velocity.** Composites of density potential temperature perturbation at 25 m AGL

(shading), water vapor mixing ratio perturbation at 25 m AGL (contours every 0.2 g kg$^{-1}$), and vertical velocity at 50 m AGL (black stippling for values greater than 0.1 m s$^{-1}$) centered over convection initiation locations for each simulation (rows) and density current category (rows). The title of each panel indicates the percentage of initiations in the given category for the given simulation.

We now use $I_{org}$ to quantify the impact of heterogeneous vegetation and cold pools on convection initiation locations (Figure 4a). Convective updrafts are close to randomly distributed with homogeneous vegetation, as the mean $I_{org}$ for HMGN and HMGN_EOFF are both ~0.5 (Figure 4a). The individual effect of cold pools ($F_{CP}$) on $I_{org}$ is small compared to the individual effect of vegetation heterogeneity ($F_{VEG}$) and the synergistic interactions between vegetation patterns and cold pools ($F_{VEG\_CP}$) (Figure 4c), demonstrating that cold pools do not have a large impact on the randomness of convection initiation locations over homogeneous vegetation (Figure 4a; Figure 2e,f). Heterogeneous vegetation acts to aggregate convection initiation locations, since $F_{VEG}$ is positive for mean $I_{org}$ (Figure 4c), consistent with our previous finding that convection initiation is less frequent over the deforested regions of CTRL and EOFF than the forests. The synergistic interactions between heterogeneous vegetation and cold pools act to disaggregate convection initiation locations, since $F_{VEG\_CP}$ is negative for $I_{org}$ (Figure 4c), consistent with cold pools propagating from the forest into the deforested regions and subsequently initiating convection in the deforested region of CTRL but not in EOFF. Thus, the effects of cold pools on convection initiation patterns are only apparent when vegetation is heterogeneous. Each distribution shown in Figure 4a is significantly different from the other distributions at the p<0.001 level.

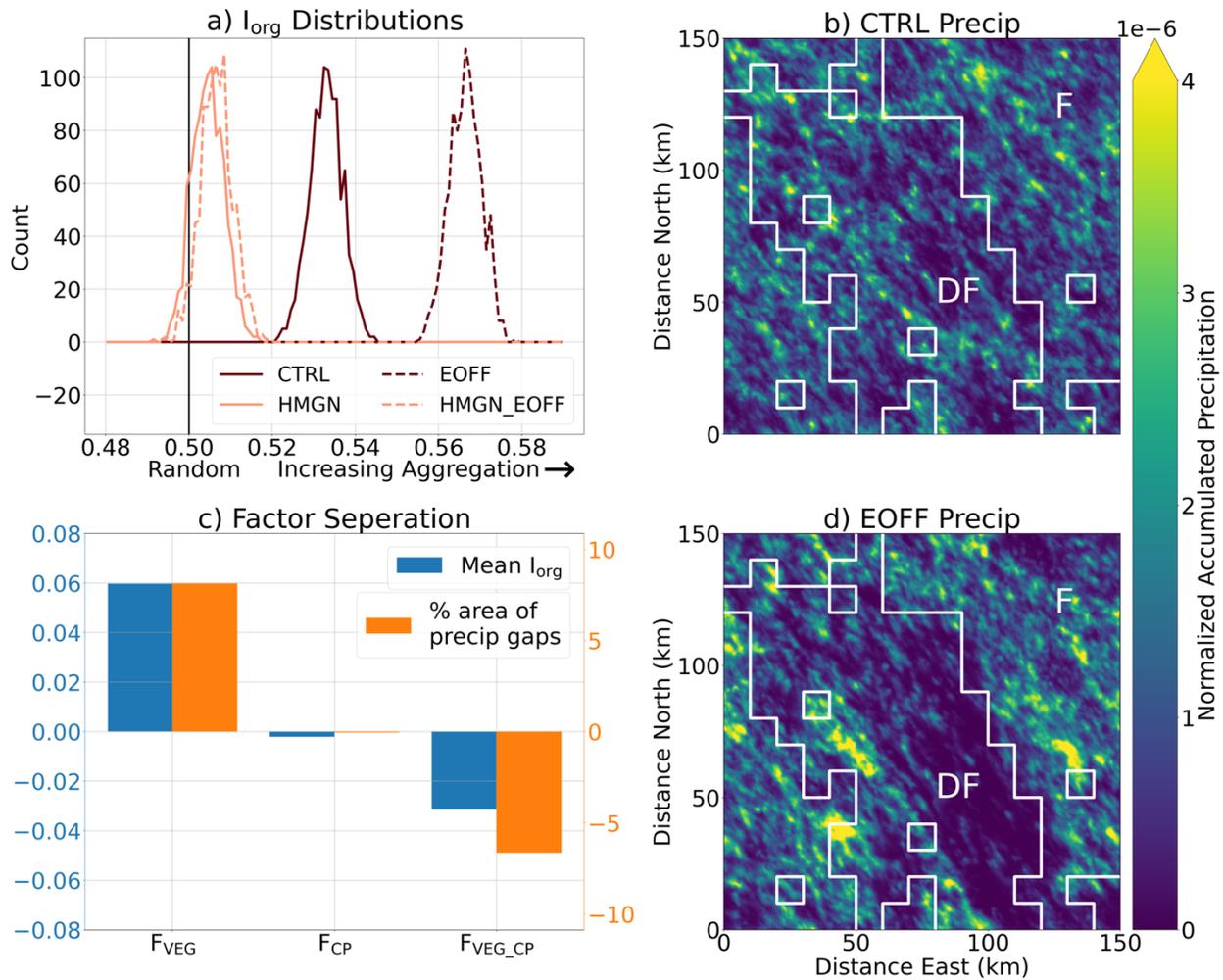

Figure 4: **Cold pools disaggregate convection initiation locations and accumulated precipitation over heterogeneous vegetation.** (a) Distributions of $I_{org}$ of convection initiation locations for each simulation. (b,d) Normalized accumulated precipitation over the five simulated days (shading) for (b) CTRL and (d) EOFF where text and white contour indicate forested and deforested areas. (c) Factor separation analysis of mean $I_{org}$ values and percentage area of precipitation gaps.

Precipitation is impacted by changes in convection initiation driven by heterogeneous vegetation and cold pools. Figure 4b (3d) shows precipitation in CTRL (EOFF), normalized by the domain mean precipitation to account for the differences in evaporation processes between CTRL and EOFF. The downwind (southeast) portion of the deforested region of EOFF receives almost no precipitation (Figure 4d), while such a pattern is less evident in CTRL (Figure 4b). These precipitation patterns follow from patterns in convection initiation and the generally southeastward motion of convective updrafts (Figure 1b, Figure 2). We define "precipitation gaps" as areas of the domain receiving <1 mm of precipitation. We determine the total area of precipitation gaps in each simulation and conduct factor separation analysis on this metric in an analogous manner as it is conducted for $I_{org}$ (Figure 4c). The factors for this precipitation gap metric have similar relative magnitudes as the factors for mean $I_{org}$, thereby demonstrating that aggregating (disaggregating) convective initiation locations causes precipitation gaps to become larger (smaller).

**4 Summary and Conclusions**

This study aimed to investigate the individual and synergistic impacts of surface heterogeneity and cold pools on the initiation of tropical continental convective storms. We find that greater surface roughness and lower albedo lead to greater surface heat fluxes and virtual potential temperatures over forested compared to deforested regions (Figure 1a,c,e). This results in more frequent convection initiation over forests than deforested regions due to favorable thermodynamics and the development of solenoidal circulations near forested and deforested boundaries (Figure 2a-d). Convection initiation locations are therefore aggregated over heterogeneous vegetation (Figure 4a). However, similar to findings by Chen et al. (2020) and Ascher et al. (2024), cold pools can propagate away from convective updrafts over the forest and initiate new convection over the less thermodynamically favorable deforested region (Figure 2a-d). This mechanism partially counteracts the effects of heterogeneous vegetation and disaggregates convection initiation locations (Figure 4a,c). Without cold pools, a large gap in the precipitation field forms over the deforested region in the heterogeneous vegetation simulation, since few convective updrafts initiate in this region (Figure 2c, Figure 4d). Convection initiation is nearly random over homogeneous vegetation, regardless of whether or not cold pools are present, so the disaggregating effect of cold pools on convection initiation locations is only important for heterogeneous vegetation (Figure 2e,f, Figure 4a).

The results of this study indicate that interactions between heterogeneous surfaces and cold pools are crucial for convection initiation patterns. While our simulations are representative of convection in the Amazon rainforest, we suggest that cold pools can similarly counteract the effects of heterogeneous vegetation in other environments. Accurately parameterizing interactions between heterogeneous surfaces and cold pools may improve the ability of forecast and climate models to represent convection initiation over heterogeneous surfaces. While not tested in this study, the background wind, which was coincidentally perpendicular to the solenoidal circulations, likely plays an important role (Allouche et al., 2023; Findell & Eltahir, 2003a, 2003b) and should be tested. This study focused on scattered convection, but results may differ for organized systems, as Maybee et al. (2025) found that Sahelian MCS initiation rates are slightly increased in a simulation with suppressed cold pools. Finally, an important takeaway from this study is that, while deforestation and other land use changes may impact the distribution of precipitation in certain regions, cold pools partially counteract these changes.

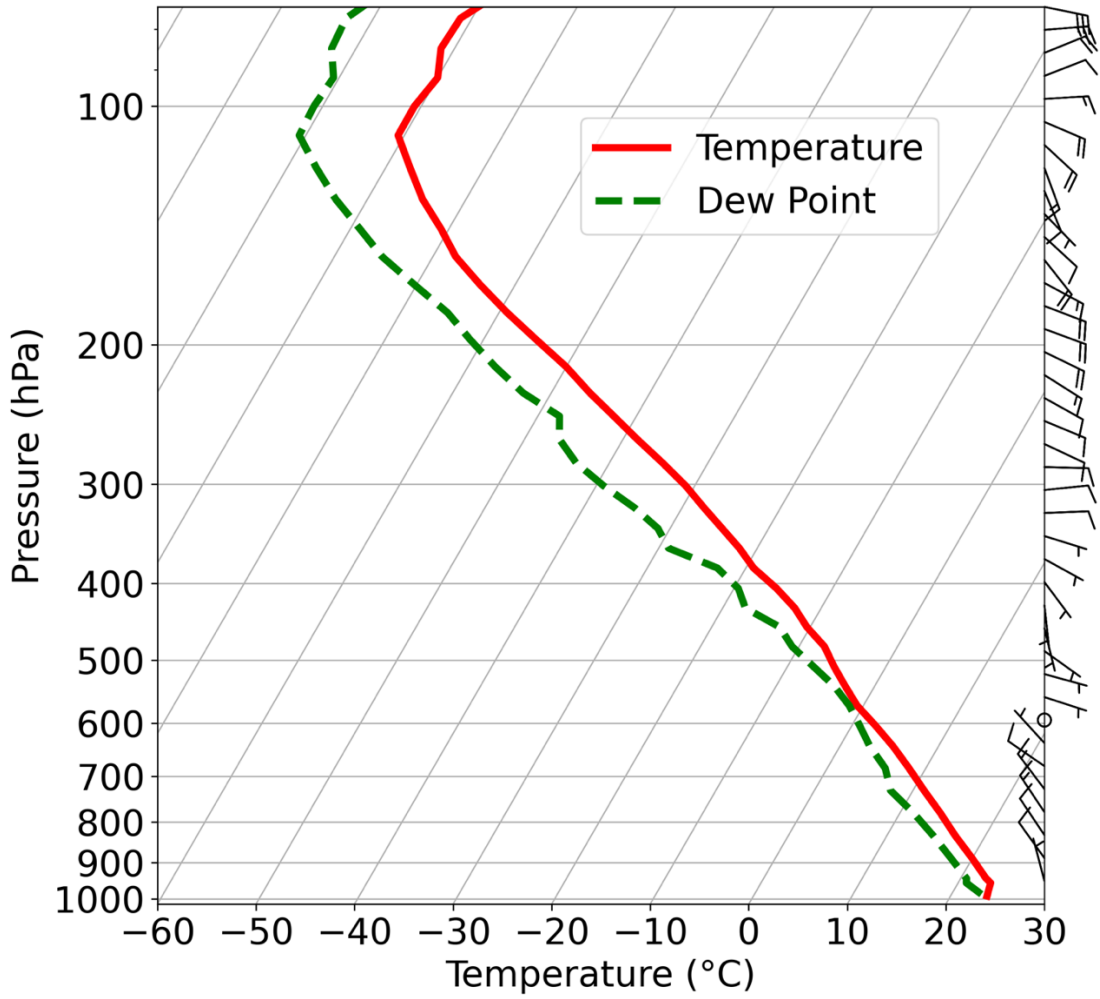

**Figure S1**: Skew T-log p diagram of the initial sounding.

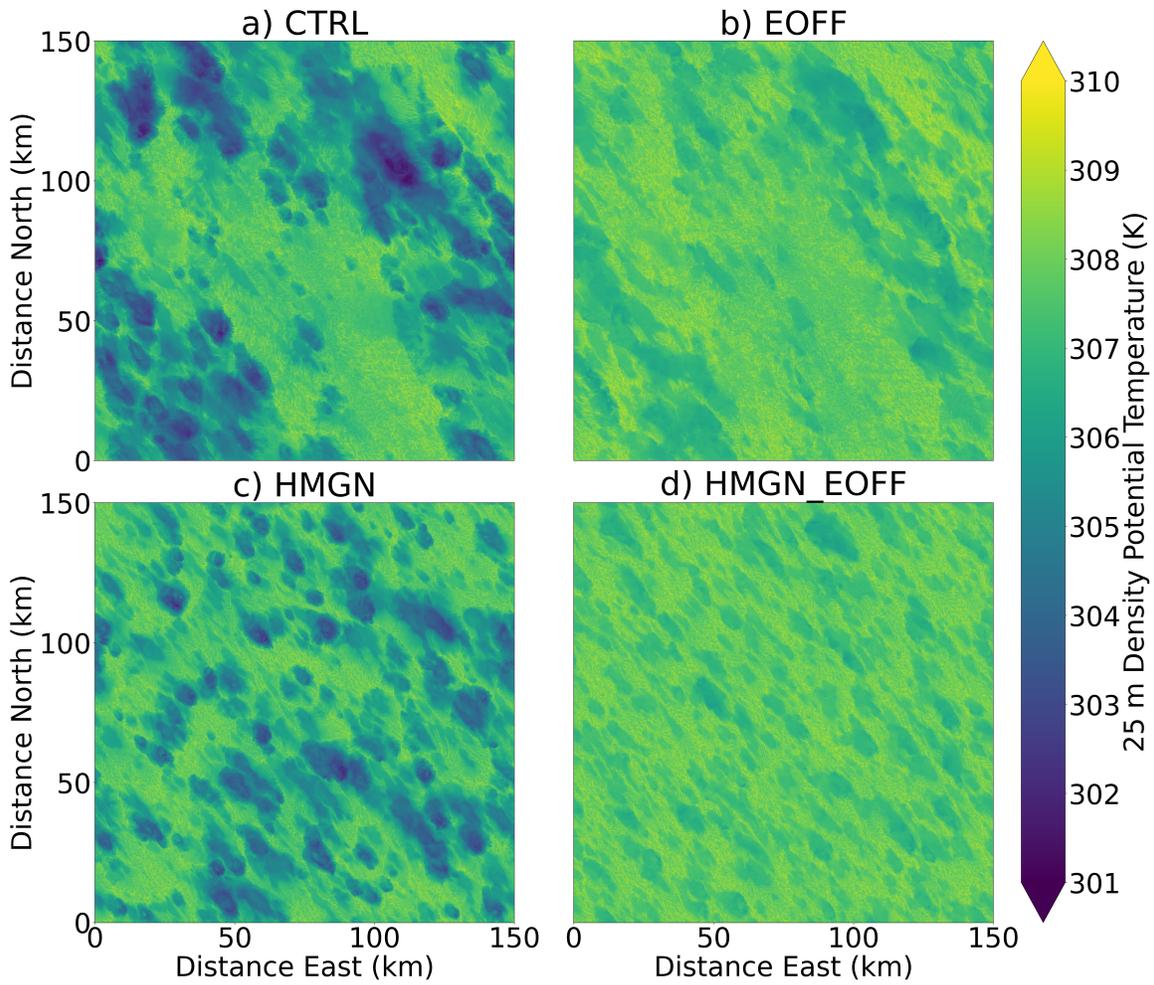

**Figure S2:** Density potential temperature at 25 m in all simulations at 1330 LT on day one.

|  | Evergreen broadleaf tree | Wooded grassland | Crop/mixed farming | HMGN and HMGN_EOFF (all points) |
|---|---|---|---|---|
| Surface roughness length (m) | 4.369 | 0.910 | 0.123 | 3.333 |
| Broadband Visible Albedo | 0.159 | 0.180 | 0.192 | 0.166 |

**Table S1:** Select surface properties.

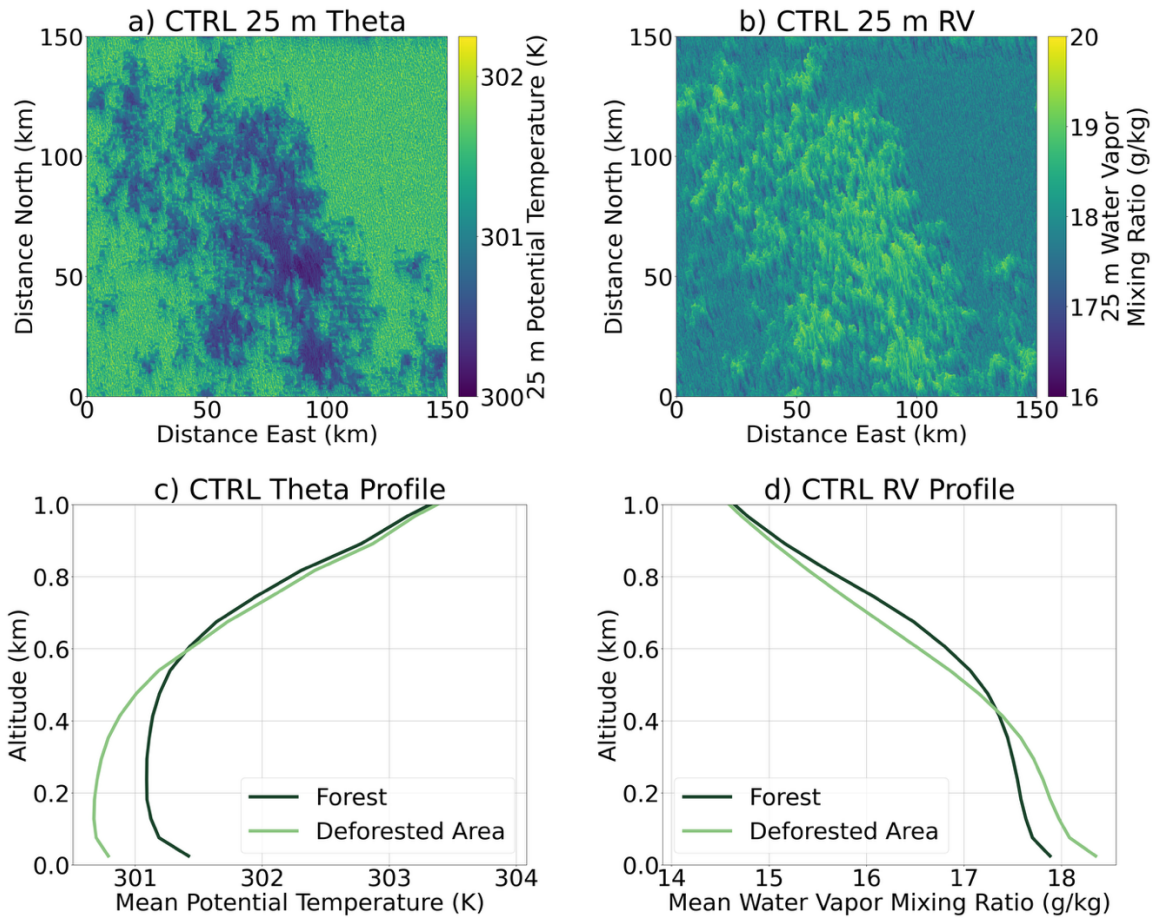

**Figure S3**: (a) Potential temperature at 25 m AGL in CTRL at 0930 LT on day one (shading). (c) Vertical profiles of mean potential temperature over the forest and deforested areas at 0930 LT on day one. (b,d) as in (a,c) but for water vapor mixing ratio.

**Appendix S1:** Factor Separation

The goal of factor separation is to isolate both the individual and synergistic effects of a set of factors on a given variable. Here, our factors are heterogenous vegetation and cold pools. We investigate the impacts of these factors on mean $I_{org}$ for each simulation. HMGN_EOFF is considered the base simulation as it has neither heterogeneous vegetation nor cold pools. The individual and synergistic impacts of heterogeneous vegetation and cold pools on $I_{org}$ can be determined by comparing $I_{org}$ in HMGN_EOFF to $I_{org}$ in the simulations with these factors. Equation S1 (identical to

Equation 1 in the main text) shows how the individual impact of heterogeneous vegetation ($F_{VEG}$) is computed:

$$F_{VEG} = I_{org,mean,EOFF} - I_{org,mean,HMGN\_EOFF} \quad \text{(S1)}.$$

Equation 1 is defined as such for the following reasons. The only difference between EOFF and HMGN_EOFF is the presence or absence of heterogeneous vegetation. Hence, any difference in mean $I_{org}$ between EOFF and HMGN_EOFF must be due to heterogeneous vegetation. $F_{VEG}$ is referred to as an individual impact, because it only considers the impact of one factor (heterogeneous vegetation).

Similarly, the only difference between HMGN and HMGN_EOFF is the presence of cold pools in HMGN and the absence of cold pools in HMGN_EOFF. Any difference in mean $I_{org}$ between HMGN and HMGN_EOFF must be due to cold pools. Hence, the individual impact of cold pools ($F_{CP}$) is calculated using Equation S2 (identical to Equation 2 in the main text):

$$F_{CP} = I_{org,mean,HMGN} - I_{org,mean,HMGN\_EOFF} \quad \text{(S2)}.$$

Now, there may be interactions between heterogeneous vegetation and cold pools which are not observed unless both of these factors are present. The impact of such synergistic interactions ($F_{VEG\_CP}$) on mean $I_{org}$ can be calculated from the difference between mean $I_{org}$ in CTRL and HMGN_EOFF, while also removing the individual effects of vegetation heterogeneity ($F_{VEG}$) and cold pools ($F_{CP}$). This is shown in Equation S3 (identical to Equation 3 in the main text):

$$F_{VEG\_CP} = I_{org,mean,CTRL} - I_{org,mean,HMGN\_EOFF} - F_{VEG} - F_{CP} = I_{org,mean,CTRL} - (I_{org,mean,HMGN} + I_{org,mean,EOFF}) + I_{org,mean,HMGN\_EOFF} \quad \text{(S3)}.$$

To understand why Equation S3 is formulated as such, consider a scenario where $F_{VEG\_CP}$ is 0. In this case, Equation S3 can be rewritten as

$$I_{org,mean,CTRL} - I_{org,mean,HMGN\_EOFF} = F_{VEG} + F_{CP} \quad \text{(S4)}.$$

In other words, if $F_{VEG\_CP}$ is 0, then the difference in mean $I_{org}$ between CTRL and HMGN_EOFF can entirely be explained by the individual (and linearly additive) impact of cold pools and the individual impact of heterogeneous vegetation. Conversely, if $F_{VEG\_CP}$ is not 0, then the two individual impacts cannot fully explain the difference in mean $I_{org}$ between CTRL and HMGN_EOFF. In this case, an interaction between cold pools and heterogenous vegetation is also influencing mean $I_{org}$, and this interaction is quantified by $F_{VEG\_CP}$."

**Appendix S2:** Categorizing Convective Updraft Initiations

Convective updraft initiations are categorized into three groups ("no/weak density currents", "single density current", and "colliding density currents") based on the presence of density currents (cold pools and/or solenoidal circulations) near each initiating convective updraft. We focus on the broader category of density currents, as cold pools and solenoidal circulations have a similar appearance and are difficult to distinguish from each other without manual analysis. The density potential temperature at 25 m ($\theta_{\rho,25m}$) around each convective initiation location is analyzed. The fields of $\theta_{\rho,25m}$ are rotated such that the coldest areas, and hence density currents, will be to the left of each convection initiation location. A temperature threshold is applied to the rotated field of $\theta_{\rho,25m}$ to determine which group ("no/weak density currents", "single density current", or "colliding density currents") each convective initiation belongs to. Composites are created by taking mean values for each group and each simulation. This procedure is described in detail below.

For each convective updraft initiation, $\theta_{\rho,25m}$ over a 6.75 x 6.75 km area centered at the location and time of initiation is considered. This area is evenly divided into 9 smaller boxes, each 2.25 x 2.25 km in size (see the final paragraph of this appendix for a discussion of the impacts of changing the size of these boxes). Figure S4a shows an example 6.75 x 6.75 km area divided into 9 numbered boxes of 2.25 x 2.25 km in size. Box 5 is in the center of the area considered and contains the location where the given convective updraft initiated. Mean values of $\theta_{\rho,25m}$ are calculated for both box 5 and box 4, which is immediately to the left of box 5. The field of $\theta_{\rho,25m}$ is then rotated counterclockwise in 9° increments without the boxes being rotated. In other words, the shading shown in Figure S4a is rotated while the white boxes remain fixed. After each rotation, mean values of $\theta_{\rho,25m}$ are recalculated for both box 5 and box 4. These values will differ for each rotation angle, as the grid of $\theta_{\rho,25m}$ is rotated relative the boxes. After testing each 9° increment between 0° and 360°, the final rotation chosen is the one which maximizes the difference in mean $\theta_{\rho,25m}$ between box 5 and box 4. Figure S4b shows the same area as Figure S4a, but after the final rotation. The 9° increments are used since a smaller increment increases computational expense without meaningfully changing results.

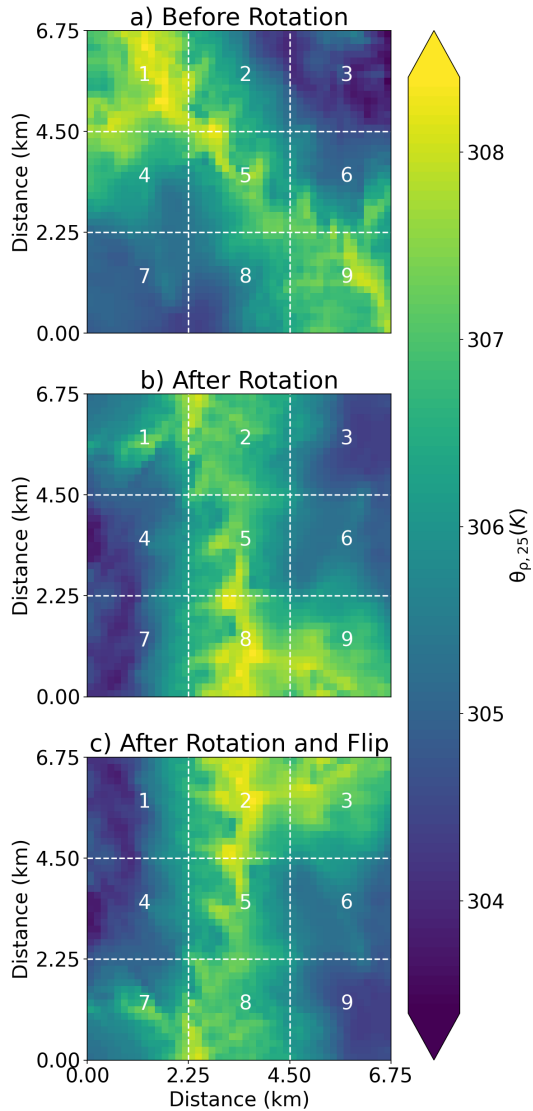

**Figure S4**: (a) $\theta_{\rho,25m}$ (shading) over an example 6.75 x 6.75 km area, boxes are indicated with white numbers and dashed lines. (b) as in (a) but after being rotated to maximize the difference in mean $\theta_{\rho,25m}$ between box 5 and box 4, therefore positioning the primary density current in box 4. (c) as in (b) but after being flipped to move the second density current from box 3 to box 9.

Now that the final rotation has been determined, if the difference in mean $\theta_{\rho,25m}$ between box 5 and box 4 is less than 0.75 K, then this convective updraft initiation is categorized as "no/weak density currents". The threshold of 0.75 K is chosen because it categorizes ~90% of convective updraft initiations as "no/weak density currents" in HMGN_EOFF. HMGN_EOFF has homogeneous vegetation and suppressed cold pools, so density currents in this simulation should predominantly be weak. Since the final rotation angle chosen is the one which maximizes the difference in mean $\theta_{\rho,25m}$ between box 5 and box 4, we do not need to check for density currents in the other boxes, as they will also fail the 0.75 K threshold.

On the other hand, if the difference in mean $\theta_{\rho,25m}$ between box 5 and box 4 is greater than 0.75 K, a density current is present in box 4. We next determine if this density current initiated the given convective updraft on its own or by colliding with another density current. If the difference in mean $\theta_{\rho,25m}$ between box 5 and box 3, box 6, and/or box 9 is greater than 0.75 K, then there is a second density current colliding with the density current in box 4, and this convective updraft initiation is categorized as "colliding density currents". Boxes 1, 2, 7, and 8 are not checked for colliding density currents because they are adjacent to box 4. If a density current is present in box 1, 2, 7, or 8, it is difficult to determine if this is the same density current as is present in box 4, or if it is a density current colliding with the density current in box 4. If the mean $\theta_{\rho,25m}$ in box 3 is colder than the mean $\theta_{\rho,25m}$ in box 9, then the array of $\theta_{\rho,25m}$ values is flipped such that the values in box 9 and box 3 are swapped. This ensures that colliding density currents are easier to identify in Figure 3c,f,i,l as they always appear in the lower right of each panel (compare Figure S4c with Figure S4b).

The remaining convective updraft initiations, which have not been categorized as "no/weak density currents" or "colliding density currents", are categorized as "single density current". Composite values of three fields, density potential temperature perturbation at 25 m AGL, water vapor mixing ratio perturbation at 25 m AGL, and vertical velocity at 50 m AGL, are shown for each simulation and category in Figure 3. At each output time, perturbations in a given field are calculated by subtracting domain mean values (for the entire simulation domain, not just the 6.75 x 6.75 km area) of that field from the value at each grid point. For each convective updraft initiation, each field is rotated/flipped in the same manner that $\theta_{\rho,25m}$ was rotated/flipped. Composites of each field are then calculated by taking the mean across all convective updraft initiations for a given simulation and category.

The procedure described above is repeated using box sizes of 1.125 x 1.125 km and 4.5 x 4.5 km along with density current thresholds of 0.50 K and 1.00 K. While the absolute values shown in Figure 3 vary when the box size or density current threshold is changed, the trends shown in Figure 3 do not qualitatively vary with the box size or density current threshold. Similarly, results do not meaningfully differ if data from 5 minutes before each convection initiation time are used instead of data at the time of each initiation.